\documentclass[aps,amsmath,amssymb,usenatbib,nofootinbib]{revtex4-2} 

\usepackage{graphicx} 

\newcommand{\abs}[1]{\left| {#1} \right|}

\begin{document}

\title{Asymmetric Light Bending in the Equatorial Kerr Metric}

\author{Arthur B. Congdon}
\affiliation{}

\author{Savitri V. Iyer}
\affiliation{Department of Physics and Astronomy, State University of New York at Geneseo, 1 College Circle, Geneseo, New York 14454, USA}

\author{Charles R. Keeton}
\affiliation{Department of Physics and Astronomy, Rutgers, the State University of New Jersey, 136 Frelinghuysen Road, Piscataway, New Jersey 08854, USA}

\begin{abstract}
The observation of the bending of light by mass, now known as gravitational lensing, was key in establishing general relativity as one of the pillars of modern physics. In the past couple of decades, there has been increasing interest in using gravitational lensing to test general relativity beyond the weak deflection limit. Black holes and neutron stars produce the strong gravitational fields needed for such tests. For a rotating compact object, the distinction between prograde and retrograde photon trajectories becomes important. In this paper, we explore subtleties that arise in interpreting the bending angle in this context and address the origin of seemingly contradictory results in the literature. We argue that analogies that cannot be precisely quantified present a source of confusion.
\end{abstract}

\maketitle


\section{Introduction}

One of the simplest formulas to emerge from general relativity (GR) is the expression for the bending angle $\hat{\alpha}$ of a light ray that passes by an object of mass $M$ at a minimum distance $r_0$:
\begin{equation} \label{alphahat-R}
    \hat{\alpha} = \frac{4GM}{c^2 r_0} \equiv 2 \, \frac{R_S}{r_0}  \,,
\end{equation}
where $R_S = 2GM/c^2$ is the object's {\it Schwarzschild radius}.
The straightforward interpretation of Eq.~(\ref{alphahat-R}) is that the amount of bending is proportional to the mass of the deflector, and inversely proportional to the distance of closest approach of the light ray to the deflector. The appearance of $c$ and $G$ is not surprising since light and gravity are involved.
Apart from a factor of two, this result can be derived from Newton's law of universal gravitation, provided that we treat a light ray as a stream of discrete particles. However, the Newtonian derivation relies on canceling out the mass of the photon, which we know to be zero. Even so, one of the great advantages of the Newtonian picture of gravity is that light bending can be understood as just one more example of classical orbital mechanics. This formulation has led to a rich literature in both astrophysics and mathematical physics (see the books \cite{Schneider-gravlenses,
Petters-singularity,
Mollerach-Roulet,
Dodelson-glbook,
Congdon-Keeton} and references therein).

Yet explaining gravitational lensing in purely classical terms misses the essentially relativistic character of the phenomenon. Consider Eq.~(\ref{alphahat-R}), for example. In order for a Newtonian description to be physically meaningful, we would expect, at the very least,  that the relativistic expression for the bending angle would reduce to Eq.~(\ref{alphahat-R}) in some appropriate limit, say, $r_0 \gg R_S$. Instead, we find a weak-field bending angle that is twice the Newtonian prediction. Explaining this discrepancy requires us to write down the exact relativistic bending angle, and expand the result in a Taylor series. The first-order term in $R_S/r_0$ yields Eq.~(\ref{alphahat-R}). 

A fully relativistic treatment of light bending shows that, in addition to the two lensed images that arise from the Newtonian bending angle, there is an infinite sequence of fainter images (see the review by \citet{Bozza-BHrev} and references therein, especially the original discovery by \citet{Darwin-exact}). If the deflector is compact enough to have a shadow border, which can be thought of as an event horizon for light, a photon that passes close to this region without falling into it can execute any number of orbits before continuing on to the observer. For a rotating compact object such as a black hole or ultra-dense neutron star, an asymmetry develops between {\it prograde} photons, which co-rotate with the deflector, and {\it retrograde} photons, which counter-rotate. The upshot for lensing is that the bending angle for prograde and retrograde photons is not the same. Both exact and approximate expressions for the bending angle in the Kerr metric have appeared in the literature, but questions of interpretation remain. After summarizing what is known about the bending angle in the Kerr metric, we identify and seek to resolve a subtle but important discrepancy that has appeared in the literature.  

\section{Light Bending in the Equatorial Kerr Metric}
\label{sec:bending}

\begin{figure}[h]
    \centering
    \includegraphics{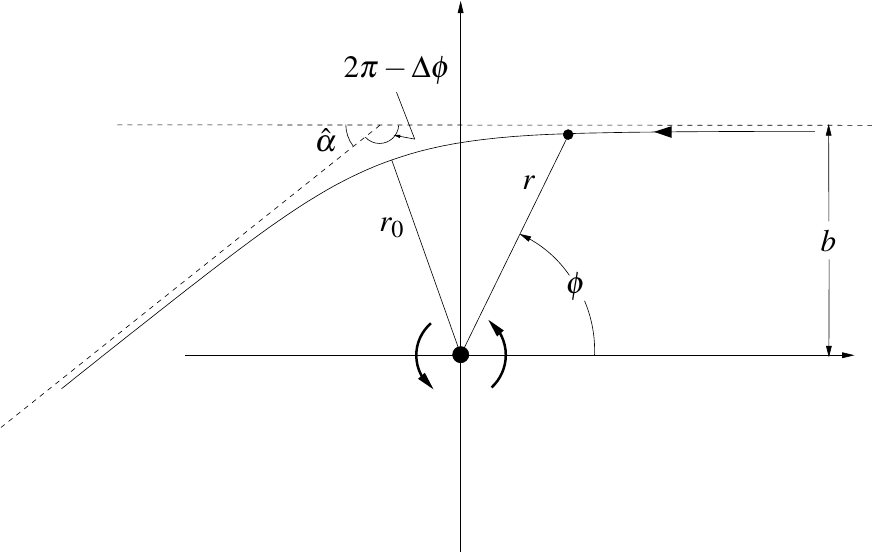}
    \caption{Lensing geometry for prograde motion in the equatorial plane of the Kerr metric. Arrows indicate the direction of propagation of the lensed photon and sense of rotation of the deflector. Retrograde motion is obtained by reflecting the photon's trajectory about the horizontal axis ($b \to -b$). Adapted from Fig. 3.1 of \citep{Congdon-Keeton}.}
    \label{fig:lensing-geometry}
\end{figure}

The spacetime around a rotating compact object is described by the Kerr metric \cite{Kerr-bh}, which is typically written in Boyer-Lindquist \cite{Boyer-Lindquist} coordinates ($t,r,\theta,\phi$), where $\theta$ is measured with respect to the axis of rotation. For simplicity, we work in the equatorial plane ($\theta=\pi/2$), where the line element takes the form
\begin{equation}
ds^2 = g_{tt}\,c^2\,dt^2 + g_{rr}\,dr^2 \ + \ g_{\phi \phi}\,d\phi^2 +  2 g_{t \phi}\, c \, dt \, d\phi.
\end{equation}
The metric coefficients are given by
\begin{subequations}
\begin{align}
&g_{tt}(r)=-\left(1-\frac{2m}{r}\right)\\
&g_{rr}(r)=\left(1-\frac{2m}{r}+\frac{a^2}{r^2}\right)^{-1}\\
&g_{\phi\phi}(r)=r^2+a^2+\frac{2m a^2}{r}\\
&g_{t\phi}(r)=-\frac{2m a}{r}
\end{align}
\end{subequations}
where $m \equiv GM/c^2$ is the mass scaled by constants to have dimensions of length. The {\em spin parameter}, which has the dimension of length, is defined by $a \equiv j/c$, where $j$ is the angular momentum per unit mass of the compact object. Without loss of generality, we take $j\geq0$. It is convenient to define the {\it dimensionless} spin parameter $\hat{a} \equiv a/m$, which is restricted to the interval $0 \leq \hat{a} \leq 1$. The Schwarzschild limit corresponds to $\hat{a}=0$, while maximal rotation corresponds to $\hat{a}=1$. The case $\hat{a}>1$, where an event horizon with some $r>0$ gives way to a naked singularity at $r=0$, is not considered in this Note.

\subsection{Equations of Motion}

The equations of motion for a particle subject to a given metric $ds$ follow from the Lagrangian $L = (ds/d \lambda)^2 / 2$, where $\lambda$ is an {\em affine} parameter. For a massive particle, we can identify $\lambda$ as the proper time. For a massless particle such as a photon, where the proper time vanishes, we can define $\lambda$ to be any parameter that specifies the spacetime coordinates everywhere along the trajectory. 

The Lagrangian for a particle in the equatorial plane of the Kerr metric is
\begin{equation}
    L(r,\dot{t},\dot{r},\dot{\phi})= \frac{1}{2} \left(c^2 g_{tt} \dot{t}^2 
    + g_{r r} \dot{r}^2 + g_{\phi \phi} \dot{\phi}^2 + 2 g_{t \phi} c \dot{t} \dot{\phi}
    \right) \,,\label{eqn:Lagrangian-equatorial-Kerr}
\end{equation}
where an overdot denotes differentiation with respect to $\lambda$. We can identify two constants of motion:
\begin{subequations}
\begin{alignat}{1}
   \varepsilon &\equiv - \frac{\partial L} {\partial \dot{t}}  
   = - c^2 g_{tt} \dot{t} - c g_{t \phi} \dot{\phi} \\
   \ell &\equiv \frac{\partial L} {\partial \dot{\phi}} 
   = c g_{t \phi} \dot{t} + g_{\phi \phi} \dot{\phi} \,\,  .
\end{alignat}
\label{eqn:vareps-ell-Kerr}%
\end{subequations}
For a massive particle, $\varepsilon$ and $\ell$ reduce to the energy per unit mass and angular momentum per unit mass, respectively, when $r \gg m$. Since these quantities do not depend on the mass of the particle, they can be straightforwardly extended to include the massless case.
It follows from our sign convention for $j$ that $\ell > 0$ for a prograde trajectory and $\ell < 0$ for a retrograde trajectory.

Solving for $\dot{t}$ and $\dot{\phi}$ in terms of $\varepsilon$ and $\ell$ yields 
\begin{equation}
\dot{t} = \frac{1}{c^2} \left( \frac{ c \ell g_{t \phi}  + \varepsilon g_{\phi \phi} }{ g_{t \phi}^2 - g_{t t} g_{\phi \phi}  }\right)
= \frac{\varepsilon}{c^2} \frac{\left[r^2 + a^2\left( 1 + \frac{2m}{r}\right) - \frac{2ma}{r}\frac{ c \ell}{\varepsilon}\right]}{\left(r^2 - 2mr +a^2\right)}
\end{equation}
and
\begin{equation}
\dot{\phi} = - \, \frac{1}{c} \left( \frac{ c \ell g_{t t}  + \varepsilon g_{t \phi} }{ g_{t \phi}^2 - g_{t t} g_{\phi \phi} } \right)
= \ell \frac{\left(1 - \frac{2m}{r} + \frac{2ma}{r} \frac{\varepsilon}{c \ell}\right)}{\left(r^2 - 2mr +a^2\right)} \, .
\label{Kerr-ang}
\end{equation}
Substituting these expressions into Eq.~(\ref{eqn:Lagrangian-equatorial-Kerr}), and setting $L=0$ along a null geodesic, we obtain
\begin{equation}
\dot{r}^2 = \frac{\varepsilon^2}{c^2} + \frac{1}{ r^2}\left(\frac{\varepsilon^2 a^2}{c^2} - \ell^2 \right)
                   + \frac{2 m}{r^3}\left(\frac{\varepsilon^2 a^2}{c^2} -  \frac{ 2 a \varepsilon \ell}{c} + \ell^2\right) \, .
\label{eqn:Kerr-rad}
\end{equation}

\subsection{Bending Angle}

In terms of the change in azimuthal angle $\phi$ as the light ray travels from the source to the observer, the bending angle is given by
\begin{align}
    \hat{\alpha} &= -\pi + \left( \int_{\rm{r_{src}}}^{r_0}  + \int_{r_0}^{{\rm r_{obs}}} \right) 
    \frac{d \phi}{dr}\,dr \nonumber \\
    &=
    -\pi + \left( - \int_{\rm{r_{src}}}^{r_0}  + \int_{r_0}^{{\rm r_{obs}}} \right)
    \abs{\frac{d \phi}{dr}} dr \, .
\end{align}
The signs of the integrals correspond to a photon traveling {\it inward} from the source at distance $r_{\rm_{src}}$ to the distance of closest approach $r_0$, and then {\it outward} from $r_0$ to the observer at $r_{\rm_{obs}}$. Since the observer and source are assumed to be far from the lens, we let $r_{\rm_{src}}, r_{\rm_{obs}} \to \infty$. (See Fig. \ref{fig:lensing-geometry}.) This leads to 
\begin{equation}
  \hat{\alpha} = -\pi + 2 \int_{r_0}^\infty 
   \abs{\frac{d \phi}{dr}} \,dr 
  = -\pi + 2 \int_{r_0}^\infty 
  \abs{\frac{\dot{\phi}}{ \dot{r}}} \,dr \,.\label{eqn:alphahat-def}
\end{equation}

To obtain the bending angle in the Kerr metric, we rewrite Eqs.~(\ref{Kerr-ang}) and (\ref{eqn:Kerr-rad}) as 
\begin{subequations}
\begin{eqnarray}
\dot{\phi} &=& \frac{\ell}{br} \left[\frac{br - 2m(b-a)}{r^2 - 2mr +a^2}\right] \\
\dot{r}^2 &=& \frac{\ell^2}{b^2 r^3}  P(r) \,, \label{eqn:phidot-rdot-b}
\end{eqnarray}
\label{eqn:phidot-rdot}
\end{subequations}
where
\begin{equation}
    P(r) \equiv r^3  - (b^2 - a^2)r + 2m (b - a)^2 \,,
\label{eqn:Pdef}
\end{equation}
and $b \equiv c \ell / \varepsilon$ is the impact parameter. Since a lensed photon follows an unbound orbit, its energy is positive ($\varepsilon>0$). Thus, $b > 0$ for a prograde trajectory ($\ell > 0$), and $b < 0$ for a retrograde trajectory ($\ell <0$).
Substituting Eqs.~(\ref{eqn:phidot-rdot}) into Eq.~(\ref{eqn:alphahat-def}) yields the Kerr bending angle,
\begin{equation}
  \hat{\alpha} = - \pi
           +  2 \int_{r_0}^\infty \sqrt{\frac{r}{P(r)}} \, 
           \frac{b r - 2 m (b - a)}{r^2 -2mr +a^2}  dr \,.\label{Kerr-bending-integral}
\end{equation}
The integral above can be expressed in terms of elliptic integrals of the third kind \citep{Iyer-Hansen-equat}. In the Schwarzschild case ($a=0$), these reduce to elliptic integrals of the first kind \citep{Darwin-exact,Iyer-Petters}.

We can express $r_0$ in terms of $a$, $b$, and $m$ by noting that $\dot{r} = 0$ when $r=r_0$. At such a {\it turning point}, $P(r)$ in Eq.~(\ref{eqn:phidot-rdot-b}) vanishes for $r \neq 0$. Depending on the value of $b$, the equation $P(r) = 0$ has either two or zero solutions with $r > 0$. First, consider the case of two solutions, denoted by $r_1$ and $r_2$ ($r_1 < r_2$). As long as $r_1$ and $r_2$ lie outside the event horizon at $r_H \equiv m+\sqrt{m^2-a^2}$, we interpret $r_1$ as the maximum distance of an outward-moving photon, and $r_2$ as the minimum distance of an inward-moving photon. In other words, $r_0 \equiv r_2$. Since $P(r) < 0$ for $r_1 < r < r_2$, $\dot{r}$ is imaginary in this region. Thus motion between $r_1$ and $r_2$ is not possible. If $P(r)=0$ has no solutions for $r > r_H$, an incoming photon will fall inside the event horizon.

The critical impact parameter $b_c$ separates the cases when $P(r)$ admits two ($\abs{b} > \abs{b_c}$) or zero ($\abs{b} < \abs{b_c}$) turning points. As $\abs{b} \to \abs{b_c}$ from above, the two positive zeros of $P(r)$ approach a common value $r_c$. Thus $P'(r_c) = 0$ when $b=b_c$. Solving for $r_c$ yields
\begin{equation}
   r_c = \sqrt{\frac{b_c^2-a^2}{3}}\,.
\label{eqn:rc-bc}
\end{equation}
To understand what $r_c$ means physically, we differentiate Eq.~(\ref{eqn:phidot-rdot-b}), and solve for $\ddot{r}$:
\begin{equation}
    \ddot{r} = \frac{\ell^2}{2 b^2} \left[ -\frac{3}{r^4} P(r) + \frac{1}{r^3}P'(r) \right] \, .
\end{equation}
Since $P(r_c) = P'(r_c) = 0$, we conclude that $\ddot{r}=0$ when $r=r_c$. Thus a photon with $b = b_c$ approaches a circular orbit at $r_c$. 
The circle of radius $r_c$ is known as the {\it shadow border} because an observer at $r>r_c$ cannot detect a photon emitted from any $r \leq r_c$. If we allow for photon trajectories that are not restricted to the equatorial plane, the shadow border becomes a two-dimensional surface.

\begin{figure}
\begin{center}
\includegraphics[width=3.0in]{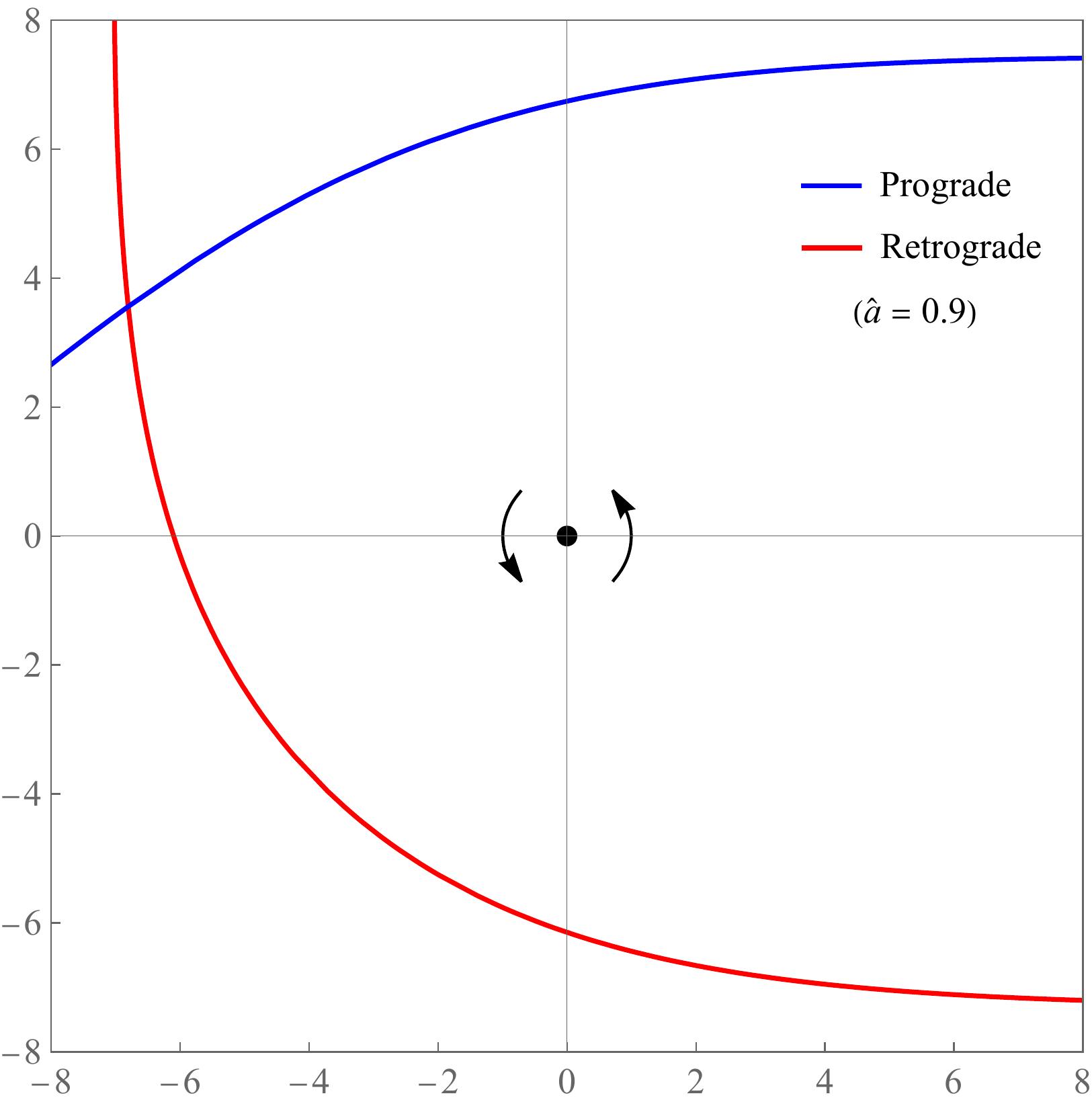}
\caption{Paths of prograde (upper curve) and retrograde (lower curve) photons for a deflector with spin parameter $\hat{a}=0.9$. The rays start out far to the right of the figure with the {\it same} unsigned impact parameter $|b|=7.5m$. The axes are labeled in units of $m$.}
\label{fig:Asymmetry} 
\end{center}
\end{figure}

To solve for $b_c$ in terms of $a$ and $m$, we substitute Eq.~(\ref{eqn:rc-bc}) into Eq.~(\ref{eqn:Pdef}), and set $P(r_c)=0$. This leads to the cubic equation
\begin{equation}
    (b_c + a)^3 = 27 m^2 (b_c - a) \,,
\end{equation}
which has a doubly degenerate solution with $b_c > 0$, and a single solution with 
$b_c < 0$. The respective critical impact parameters for prograde and retrograde motion are denoted by $b^+_c$ and $b^-_c$, and are given by 
\begin{equation}
\label{bcrit}
b^{\pm}_c = -a \pm 6m 
  \cos{\left[\frac{1}{3}\cos^{-1}{\left(\mp \frac{a}{m}\right)}\right]} \,.
\end{equation}
In the non-rotating case ($a=0$), we recover $b_c = 3m$.
In the case of maximal rotation ($a=m$), $b_{c}^{+} = 2m$ and $b_{c}^{-} = -7m$.

\section{Asymmetry between Prograde and Retrograde Motion}
\label{sec:conclusion}

It is not obvious from Eq.~(\ref{Kerr-bending-integral}), or from the representation of the bending angle in terms of elliptic integrals, that prograde and retrograde photons are deflected by different amounts. This asymmetry becomes apparent when we consider the bending angle in two asymptotic cases: the {\it weak deflection limit} (WDL) and the {\it strong deflection limit} (SDL). In the WDL, $\abs{b} \gg m$, so that the bending angle can be written as a power series in $m/\abs{b}$. Starting respectively from the elliptic integral representation of the bending angle and from Eq.~(\ref{Kerr-bending-integral}), \citet{Iyer-Hansen-arXiv} and \citet{Aazami-2011b} find 
\begin{equation}
\begin{split}
\label{AlphaKWeak}
\hat{\alpha}_{\pm}&=4\left(\frac{m}{|b|}\right)+\left(\frac{15\pi}{4}
\mp4\hat{a}\right) \left(\frac{m}{|b|}\right)^2\\
&\quad\quad+\left(\frac{128}{3}\mp 10 \pi\hat{a}+4\hat{a}^2\right) \left(\frac{m}{|b|}\right)^3 \\
&\quad\quad\quad+\left(\frac{3465\pi}{64} \mp192\hat{a}+\frac{285\pi}{16}\hat{a}^2 \mp4\hat{a}^3\right)
\left(\frac{m}{|b|}\right)^4\\
&\quad\quad\quad\quad+O\left[\left(\frac{m}{|b|}\right)^5\right] \,
\end{split}
\end{equation}
in the equatorial plane of the Kerr metric for a fixed value of $\hat{a}$. Apart from the first-order term in $m/\abs{b}$, which is independent of spin, all other terms increase (decrease) the bending angle of a retrograde (prograde) photon relative to the non-rotational case.

In the SDL, where $\abs{b} \sim m$, a power series representation of the bending angle would require one to identify a different expansion parameter. However, this would prove to be an exercise in futility, since the bending angle diverges logarithmically \citep{Ohanian-BH} as 
$\abs{b} \to \abs{b_c} \sim m$.
This same behavior is seen for the Schwarzschild bending angle \citep{Darwin-exact}. Thus, we examine the bending angle numerically in this case. Figure \ref{fig:Asymmetry} shows the trajectories of two photons with the same value of $\abs{b}$, but on opposite sides of the horizontal axis. Notice that the bending occurs mostly within a few gravitational radii of the deflector. To understand why the retrograde photon is bent more than its prograde counterpart, there is an additional asymmetry in Kerr lensing that warrants discussion. The critical radius for photon capture takes on the single value $r_c = 3m$ in the Schwarzschild metric, but has two different values, $r_{c}^{+}$ (prograde) and $r_{c}^{-}$ (retrograde) in the Kerr metric, which depend on the mass and spin of the deflector. They approach the Schwarzschild value $r_c = 3m$ as $\hat{a} \to 0$. At maximal rotation ($\hat{a} = 1$), $r_{c}^{+} = m$ and $r_{c}^{-} = 4m$. Since $r_{c}^{+} < r_{c}^{-}$, a prograde photon can get closer to the deflector than a retrograde photon without being captured. A less intuitive but more useful equivalent statement is that, for a fixed value of $\abs{b}$, a retrograde photon gets closer to its critical radius than a prograde photon gets to its critical radius. Thus, the bending angle of the retrograde photon is the larger of the two.

\begin{figure*}
\begin{center}
\includegraphics[height=6.0cm]{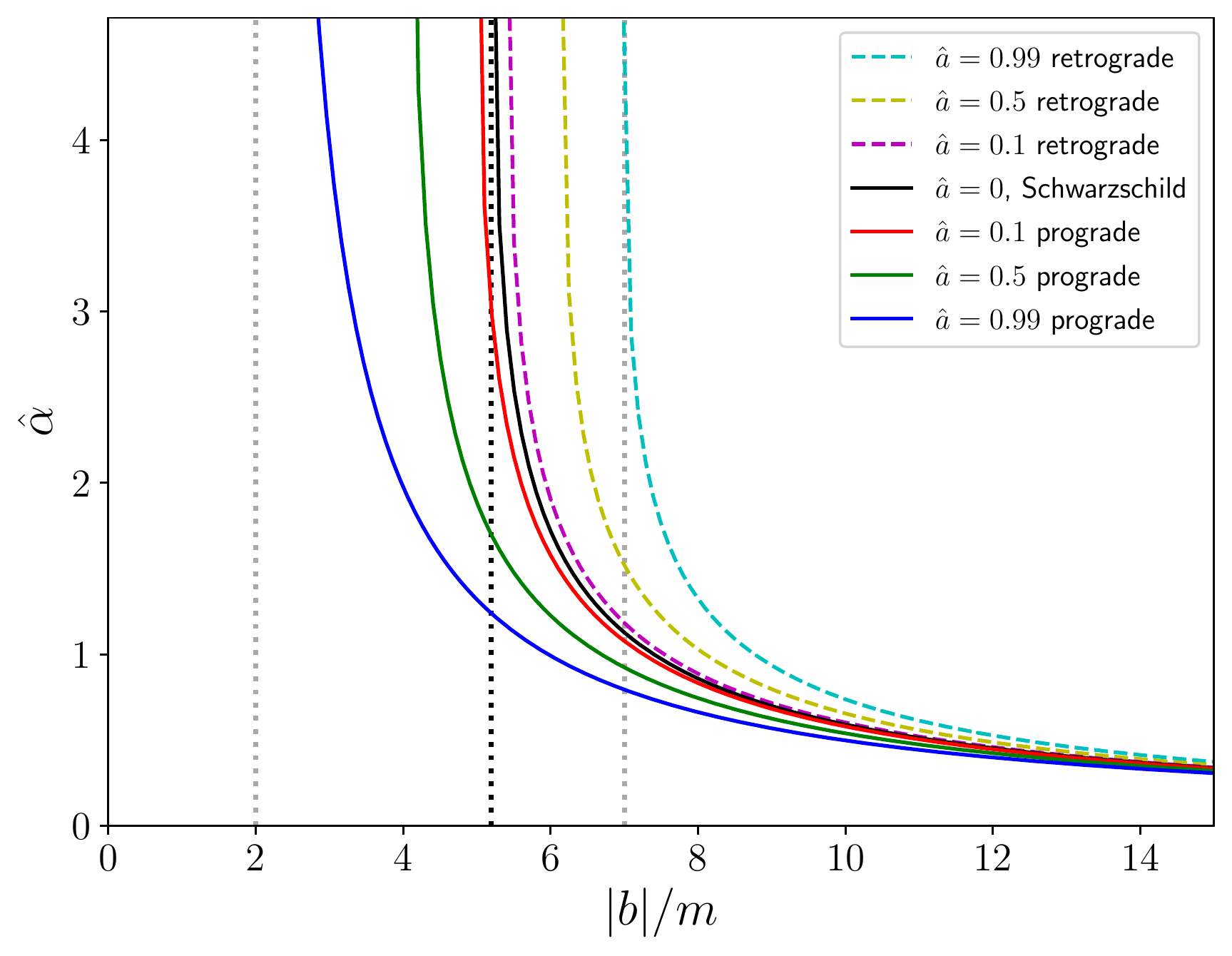}
\includegraphics[height=6.0cm]{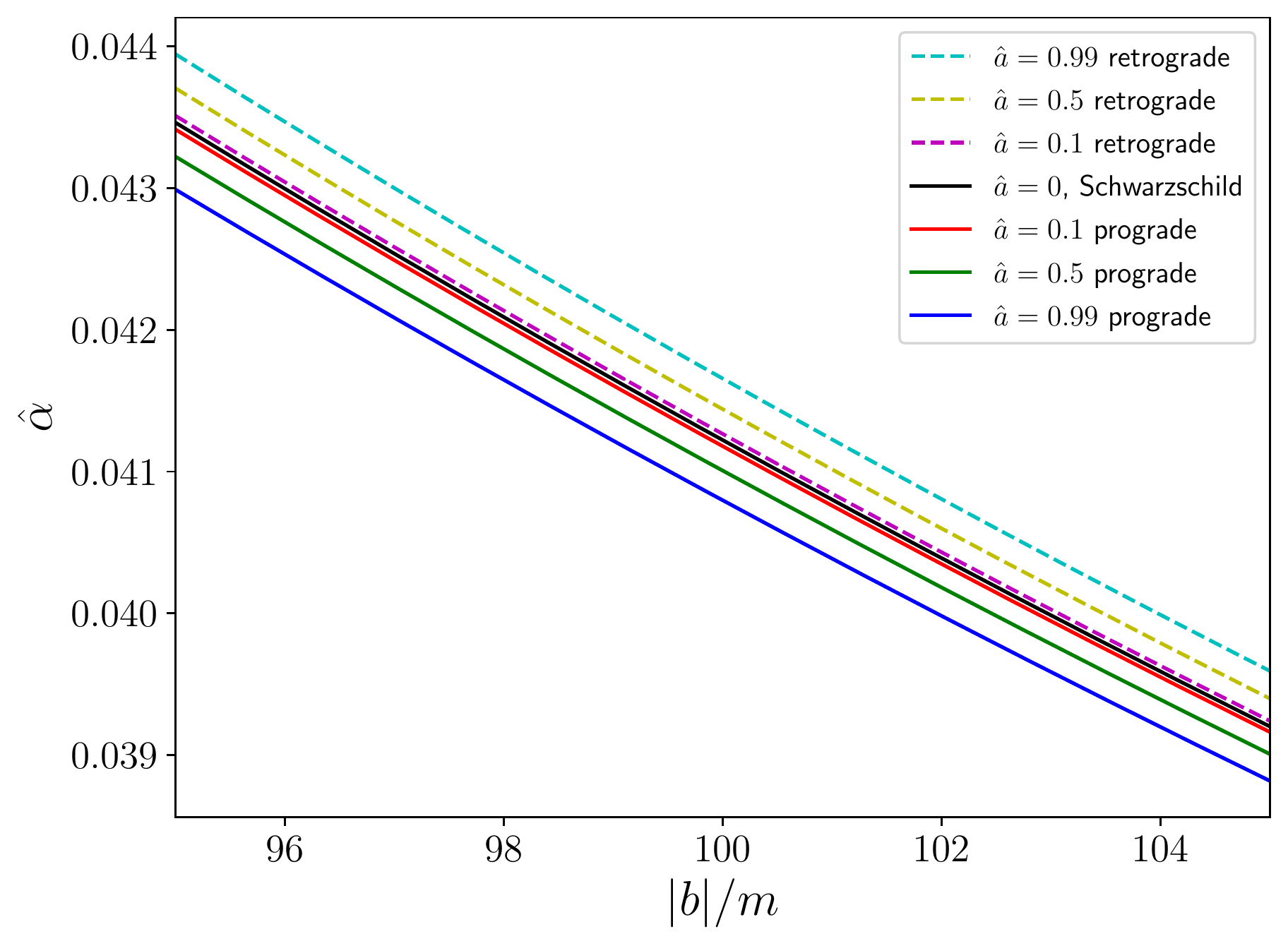}
\caption{
Exact bending angle $\hat{\alpha}$ as a function of $\abs{b}/m$ for different values of the spin parameter $\hat{a}$. For Kerr cases, prograde rays are shown as solid lines and retrograde rays are shown as dashed lines. The left panel shows the strong deflection limit; the vertical dotted lines indicate the critical impact parameter $b_c/m = 3\sqrt{3}$ for the Schwarzschild case (black) as well as the limits 
$b_c^+/m= 2$ and $\abs{b_c^-}/m=7$ that correspond to prograde and retrograde photons, respectively, for $\hat{a} = 1$ (gray). The right panel shows the weak deflection limit. (The colors and line styles are the same in both panels.)
}
\label{fig:ExactBendingAngle} 
\end{center}
\end{figure*}

For the sake of plotting the bending angle in the Kerr and Schwarzschild metrics for both prograde and retrograde motion over the full range 
$\abs{b_c^{\pm}} < \abs{b} < \infty$, \citet{Iyer-Hansen-equat} introduce the parameter $b' \equiv 1-\abs{b_{c}^{\pm}/b}$, which ranges from zero in the extreme SDL to unity in the extreme WDL. They plot the bending angle against $b'$ in their Figs.~4 and 5, finding greater deflection for prograde photons than for retrograde photons with the same value of $b^{\prime}$.
Yet this conclusion runs counter to our discussion above. This is a direct consequence of the dependence of $b^{\prime}$ on $b_c^{\pm}$, which has distinct values for prograde and retrograde motion. In other words, prograde and retrograde photons with the same value of $b^{\prime}$ have different values of $b$, and vice-versa. 
The most natural way around this difficulty is to plot the bending angle against the physical impact parameter separately for the WDL and SDL (see Fig.~\ref{fig:ExactBendingAngle}). In both limits, we see that the magnitude of the bending angle is indeed larger for retrograde photons than prograde photons, with the Schwarzschild value lying somewhere in between.

It is tempting to interpret the bending angle in the prograde and retrograde cases by means of analogy with a rotating fluid.  A retrograde photon, which moves ``upstream," has to overcome the ``gravitational current" induced by the deflector on its way to the observer, while a prograde photon, which moves ``downstream," is swept along with the current.   This results in a larger bending angle for a retrograde photon than for a prograde photon. Unfortunately, the same analogy can lead to the opposite conclusion \cite{Iyer-Hansen-equat}. \citet{Aazami-2011b}, who find a larger bending angle for retrograde photons, use their analysis of time delays to argue that retrograde photons ``spend more time" in the gravitational field of the deflector than prograde photons. The apparent tension between \citet{Iyer-Hansen-equat} and \citet{Aazami-2011b} disappears when the mathematical results of the two papers are compared, without reference to analogies or figures. Thus, analogies that are intended to explain physical results can become an accidental exercise in confirmation bias instead. Finding intuitive explanations for results in a field as non-intuitive as GR is a particularly risky undertaking.

\section*{Acknowledgments}

ABC would like to thank Ted Burkhardt, Jerod Caligiuri, Carl Droms,  Tim Jones, Allan Moser, and Erik Nordgren for their helpful comments on the manuscript.
SVI gratefully acknowledges Robert Sinesi, who prepared Figure \ref{fig:Asymmetry} and was supported by the NASA NY State Space Grant.


\bibliography{bibli.bib}

\end{document}